\documentclass [epjCONF,draft] {svjour} 
\usepackage[final]{graphicx}
\usepackage[varg]{txfonts} 
\usepackage[latin1]{inputenc}
\usepackage{color}
\usepackage{ulem}
\usepackage{booktabs}
\usepackage{units}
\usepackage[mathlines]{lineno}

\definecolor{lightblue}{rgb}{0,0.8,.8}

\def \pao {~[Pierre Auger Collaboration]}
\def \tac {~[Telescope Array Collaboration]}

\session-title{UHECR 2012 Symposium, CERN, February 2012}
\begin{document}

\title{The Energy Spectrum of Cosmic Rays at the Highest Energies}
\author{Bruce R. Dawson\inst{1} \and Ioana C. Mari\c s\inst{2}, Markus
  Roth\inst{3} \and Francesco Salamida\inst{4} \and Tareq
  Abu-Zayyad\inst{5} \and Daisuke Ikeda\inst{6} \and Dmitri
  Ivanov\inst{5,7} \and Yoshiki Tsunesada\inst{8} \and Mikhail
  I. Pravdin\inst{9} \and Artem V. Sabourov\inst{9}, for the Pierre
  Auger, Telescope Array and Yakutsk Collaborations} \institute{School
  of Chemistry \& Physics, The University of Adelaide, Adelaide, S.A.,
  Australia \and Laboratoire de Physique Nucl\'{e}aire et de Hautes
  Energies (LPNHE), Universit\'{e}s Paris 6 et Paris 7, CNRS-IN2P3,
  Paris, France \and Karlsruhe Institute of Technology - Campus North
  - Institut f\"{u}r Kernphysik, Karlsruhe, Germany \and Institut de
  Physique Nucl\'{e}aire d'Orsay (IPNO), Universit\'{e} Paris 11, CNRS-IN2P3,
  Paris, France \and University of Utah, Department of Physics and
  High Energy Astrophysics Institute, Salt Lake City, Utah, USA \and
  Institute for Cosmic Ray Research, University of Tokyo, Kashiwa,
  Japan \and Rutgers-The State University of New Jersey, Department of
  Physics and Astronomy, Piscataway, New Jersey, USA \and Graduate
  School and Engineering, Tokyo Institute of Technology, Tokyo, Japan
  \and Yu.G. Shafer Institute of Cosmophysical Research and Aeronomy,
  Yakutsk, Russia}

\abstract{One of several working groups established for this workshop
  was charged with examining results and methods associated with the
  UHECR energy spectrum.  We summarize the results of our discussions,
  which include a better understanding of the analysis choices made by
  groups and their motivation.  We find that the energy spectra
  determined by the larger experiments are consistent in normalization
  and shape after energy scaling factors are applied.  Those scaling
  factors are within systematic uncertainties in the energy scale, and
  we discuss future work aimed at reducing these systematics.
} 

\maketitle

\section{Introduction}

The energy spectrum working group (WG) was established approximately
two months in advance of this workshop with membership from the Pierre
Auger Observatory, the Telescope Array (TA) and the Yakutsk experiment.  In
addition, some of our members had been part of the HiRes
collaboration, and there were strong links with the AGASA experiment.
Our charge was to assess the current status of our knowledge of the
UHECR energy spectrum, and to understand more clearly the analysis
methods employed by the experiments, and their motivations.

All of the currently operational experiments (Auger, TA, Yakutsk)
recognize the advantages of using an optical technique for calibrating
the energy parameter of their surface detector arrays.  Auger and TA
use fluorescence light, while Yakutsk uses Cherenkov light.  Apart
from this common base, there are some significant differences in
analysis methods.  

We were pleased to confirm that there are no major inconsistencies
between the energy spectra obtained by Auger, HiRes and TA.  Future
work will endeavor to understand the differences that do exist, and to
take advantage of worldwide experience to hone our methods.


\section{Scope of the Working Group Discussions}

At the beginning of the WG a framework was determined for its work.

\begin{enumerate}
\item Investigate the level of agreement in the normalization and shape of
the UHECR energy spectrum determined by the groups;

\item Understand the methods for determining aperture and exposure,
  and compare the current exposures;

\item Understand the methods and motivations for determining energy
  via the surface detector (SD) of the experiments, including how
  zenith angle dependence (attenuation) is handled;

\item Discuss methods for fluorescence detector (FD)/optical
  calibration of the SD energy scale, and examine how these
  calibrations differ from those determined via simulation;

\item Examine the systematic uncertainties in the energy scale of each
  experiment, and discuss how these may be reduced in the future.  If
  some systematics are common to the experiments, can we agree on
  using a particular method or measurement? One example of this is the
  fluorescence yield description.

\end{enumerate} 
During the two month working period, the WG communicated via a Wiki
page, facilitating a collection of documents.  The WG also held four
productive Skype sessions which allowed an easy exchange of information.

\begin{figure}
\center{\includegraphics[width=0.8\textwidth]{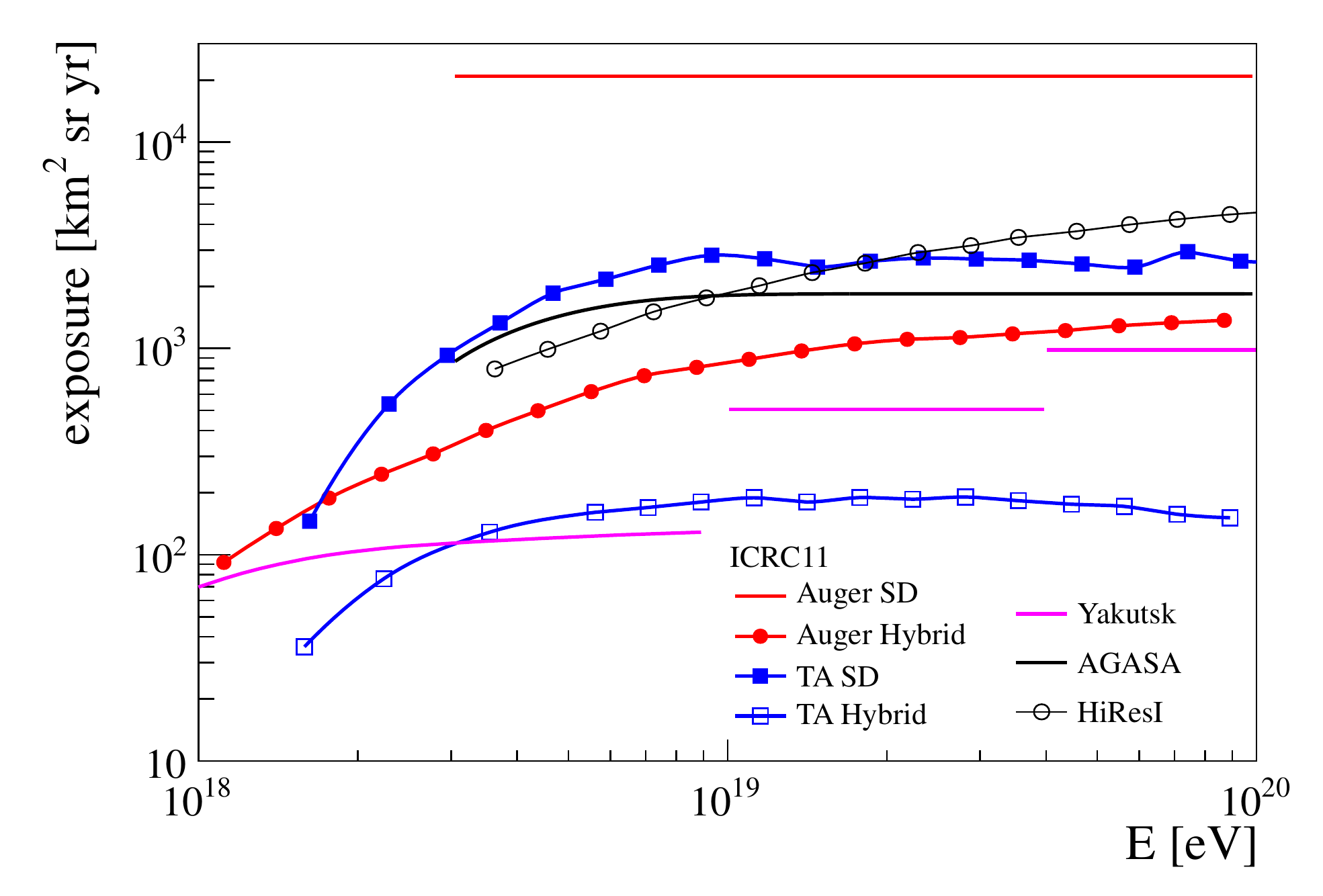}}
\caption{Exposures of the experiments at the time of the ICRC 2011
  conference for Auger SD~\cite{auger}, Auger
  Hybrid~\cite{auger}, TA SD~\cite{TASD}, TA
  Hybrid~\cite{TAHybrid}, Yakutsk SD~\cite{Yakutskspectrum}, AGASA~\cite{agasaexposure} and
  HiRes I~\cite{hiresIexposure}.  After M. Unger~\cite{Unger}.}
\label{fig:exposure}
\end{figure}

\begin{figure}
\center{\includegraphics[width=0.8\textwidth]{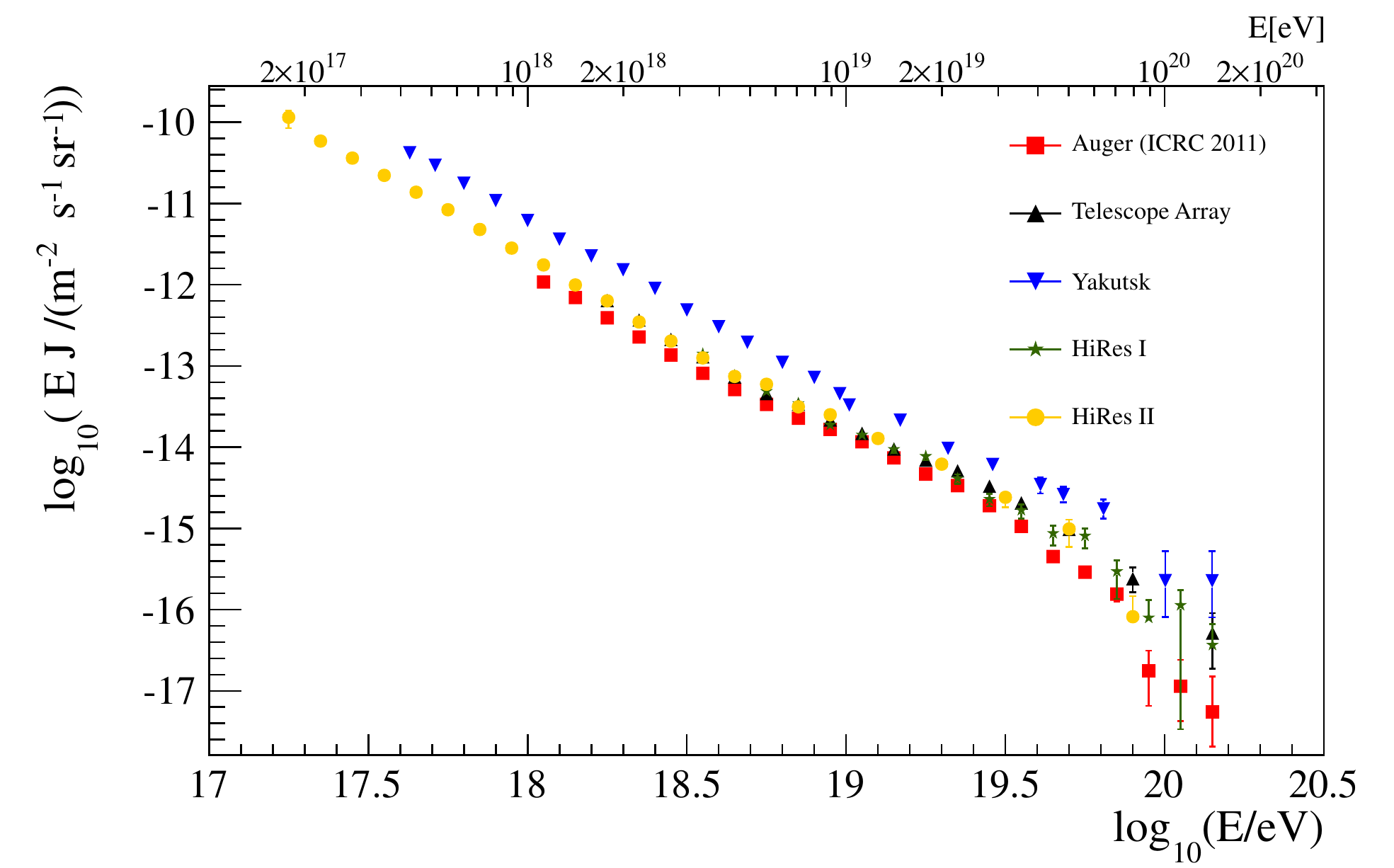}}
\caption{Published energy spectra, with the flux multiplied by $E$,
  for Auger (combined Hybrid/SD)~\cite{auger}, TA
  SD~\cite{TASD}, Yakutsk SD~\cite{Yakutskspectrum}, HiRes I~\cite{HRSpectrum}, and HiRes II~\cite{HRSpectrum}.}
\label{fig:current}
\end{figure}

\section{Comparing Energy Spectrum Measurements}

The exposures (in km$^2$sr\,yr) of some past and current experiments
are shown in Figure~\ref{fig:exposure} as a function of energy. In the
case of fluorescence detector spectra (Auger Hybrid, TA Hybrid and
HiRes I), the exposure must be calculated using detector simulations
and knowledge of the detector live time and atmospheric conditions.
Such exposures generally increase with energy, though in the case of
TA Hybrid the exposure saturates because it is limited by the
aperture of the surface detector array.
Surface detector
exposures are generally more robust, because they are calculated by
integrating the geometrical aperture of the surface array over the
live time of the experiment.

The published energy spectra are shown in Figure~\ref{fig:current}.
This particular representation, with the flux axis multiplied by a
single power of energy, is related to the actual measurement where a
number of air showers is detected in a particular logarithmic energy
interval,
($\frac{{d}\! N}{{d}\! \log_{10}E}= \frac{{d}\! N}{{d}\! E}\;E \ln
10 \propto JE $). Systematic differences are apparent.   However all experiments shown here detect the spectral
flattening at the ankle.  With the exception
of Yakutsk (due to lower exposure), all the experiments observe a
suppression, perhaps the GZK cut-off, at the highest energies.

A traditional way to characterize spectra is via a broken power-law
fit with two break-points at the ankle energy $E_A$ and the beginning
of the suppression at energy $E_S$.  Around these energies the
spectrum has a power law form $E^{-\gamma}$ with indices $\gamma_1$
below the ankle, $\gamma_2$ between the ankle and the suppression, and
$\gamma_3$ above the suppression.  The parameters are shown in
Table~\ref{tab:features}.  The fits were done by the various
collaborations (references listed in the caption) except for Yakutsk,
where the fit was done by the WG.  The HiRes fits were done after
combining the mono spectra from HiRes\ I\ \&\ II~\cite{HRSpectrum}.

We have included results from AGASA in this table, given its
importance in our field.  As is well known, the AGASA energy
spectrum~\cite{agasaspectrum} did not show evidence for a high-energy
suppression, in contrast with HiRes and Auger.  This was part of the
motivation behind the formation of the Telescope Array experiment, a
combination of fluorescence detectors and a plastic scintillator
surface detector.

\begin{table}[t] 
\begin{center}
\begin{tabular}{llllll}
\toprule
 & $\gamma_1$ & $\gamma_2$ & $\gamma_3$ & $\log_{10} E_A$ &  $\log_{10} E_S$ \\
\midrule
AGASA &$3.16 \pm 0.08$ & $2.78 \pm 0.3$ & - & $19.01$ & \\   
Yakutsk &$3.29 \pm 0.17$ & $2.74 \pm 0.20$ & - & $19.01 \pm 0.01$ & - \\
HiRes &$3.25 \pm 0.01$ & $2.81 \pm 0.03$ & $5.1 \pm 0.7$ & $18.65 \pm 0.05$ & $19.75 \pm 0.04$ \\
Auger & $3.27 \pm 0.02$ & $2.68 \pm 0.01$ & $4.2 \pm 0.1$ & $18.61 \pm 0.01$ & $19.41 \pm 0.02$ \\
TA &$3.33 \pm 0.04$ & $2.68 \pm 0.04$ & $4.2 \pm 0.7$ & $18.69 \pm 0.03$ & $19.68 \pm 0.09$ \\
\bottomrule
\end{tabular} 
\caption{Results of a triple power-law fit to the UHECR spectrum.  Fit results 
are from AGASA~\cite{agasaspectrum}, the combined HiRes\ I\ \&\ II mono spectra~\cite{HRSpectrum}, Auger (Hybrid + SD)~\cite{auger} and Telescope Array (SD)~\cite{TASD}. The fit to the Yakutsk SD spectrum~\cite{Yakutskspectrum} was done by the WG.}
\label{tab:features}
\end{center}
\end{table}

\begin{figure}
\begin{center}
\includegraphics[scale=0.5]{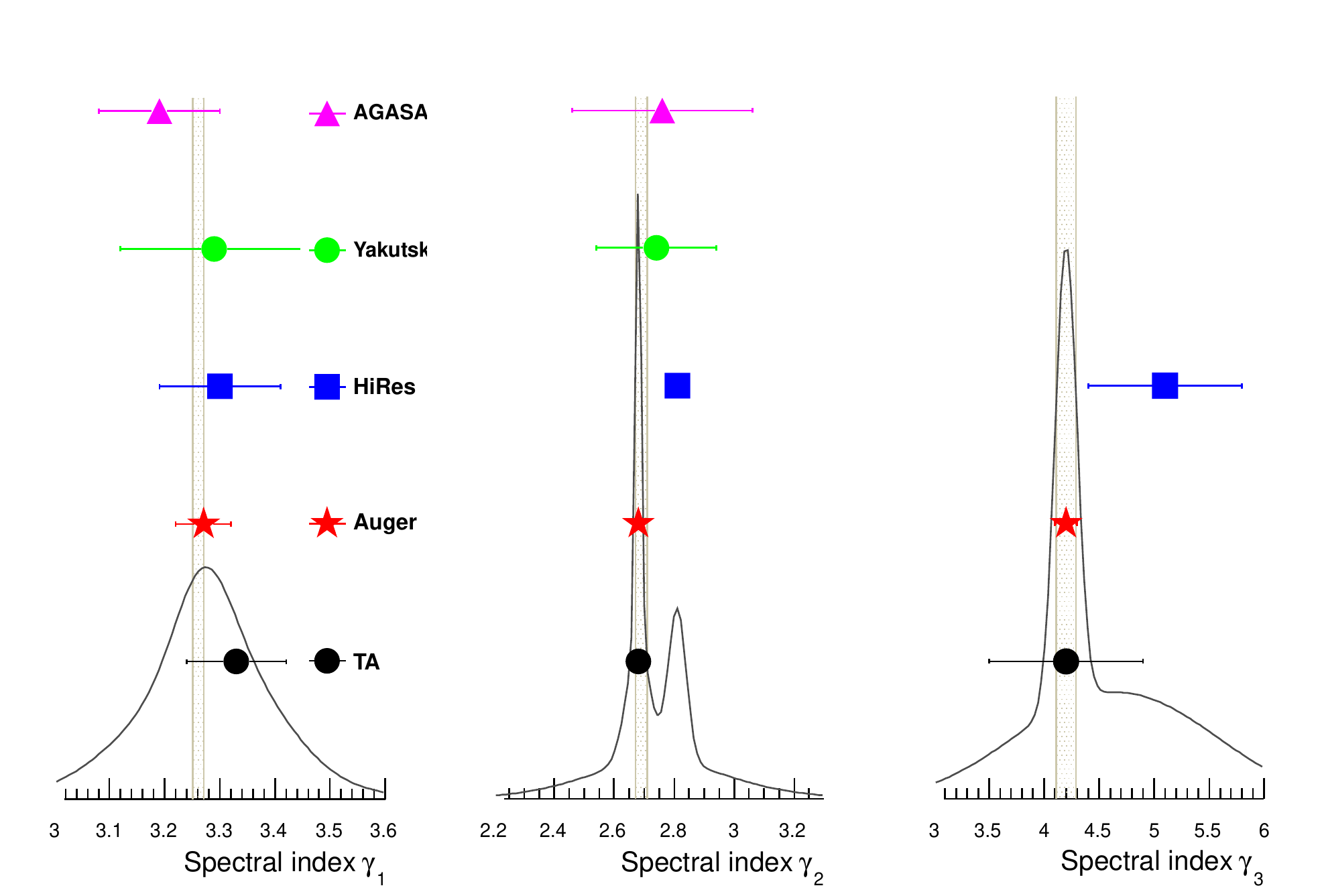}
\end{center}
\caption{Spectral indices for the triple-power law fits from
  Table~\ref{tab:features}.  The solid black line is the sum of the
  Gaussian probability distributions implied by the errors in the
  indices. The shaded bar represents the weighted mean of the
  measurements and its uncertainty.}
\label{fig:indices}

\begin{center}
\includegraphics[scale=0.5]{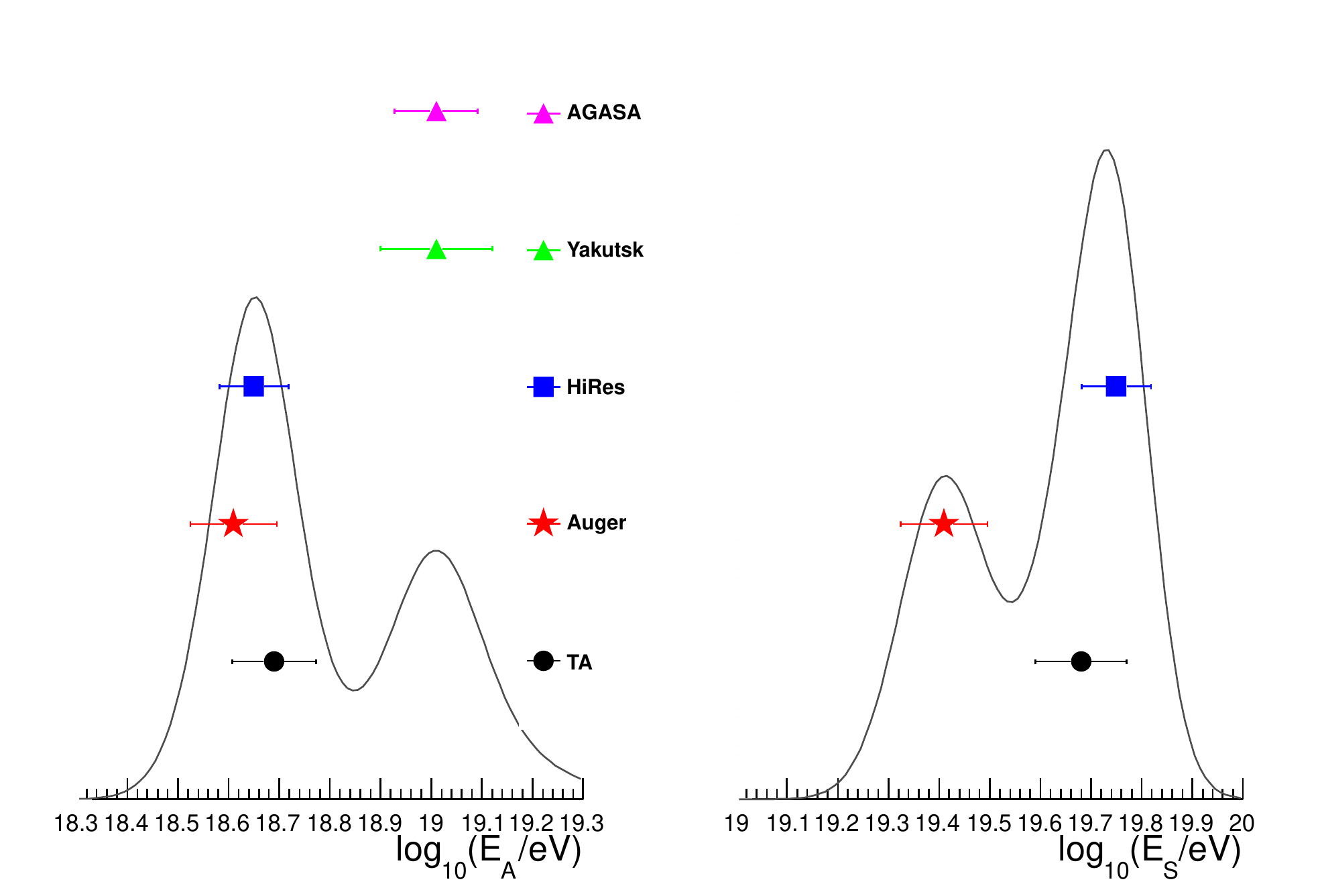}
\end{center}
\caption{Break-point energies for the triple-power law fits from
  Table~\ref{tab:features}.  The solid black line is the sum of the
  Gaussian probability distributions implied by the errors in the
  break-point energies.  Those errors are the statistical uncertainty
  folded with the systematic uncertainty in the energy scale.}
\label{fig:BreakPoints0}
\end{figure}

These fits results are displayed graphically in
Figure~\ref{fig:indices} and Figure~\ref{fig:BreakPoints0}.  We use
the quoted uncertainties of the fits to produce a probability
distribution for each parameter.  We observe that there is general
agreement on the values of spectral indices, with the possible exception of
$\gamma_2$.  There is more disagreement on the positions of the break-points in 
Figure~\ref{fig:BreakPoints0}.  

The energies relating to spectral features can be affected by
experimental energy resolution and by systematic energy
uncertainties.  We discuss each of these in turn.

\subsection{Treatment of Detector Resolution}


The differential cosmic ray intensity in a given energy interval
is determined experimentally by $n_i = N_{\rm obs}^i/A_0 T$, where
$A_0$ is the geometrical aperture of the detector (area $\times$ solid
angle), $T$ is the observation time (the exposure is defined as ${\cal
  E} = A_0 T$), and $N_{\rm obs}^i$ is the number of observed cosmic
ray events in the energy bin $i$ in the duration $T$.  Because of the
energy resolution determined by the detector performance and the
analysis algorithm, however, bin-to-bin migrations affect the
number of events in the energy bins and hence the energy spectrum.

The effect of energy resolution on the spectrum is treated slightly
differently by the existing experiments.  As was the practice with
HiRes, the Telescope Array experiment corrects for the resolution by
including its effects into the exposure calculation, for both the
surface detector spectrum and their hybrid spectrum. They begin by
generating Monte Carlo showers (QGSJET-II model) assuming a pure proton
composition and the spectrum determined by HiRes~\cite{HRSpectrum},
then passing the showers through a detailed detector simulation.
Comparisons of various real data distributions with those from
simulations give confidence in the simulations.  The exposure for each
energy bin is then calculated as $(N_{\rm rec} / N_{\rm th}) A_0 T$ where
$N_{\rm rec}$ is the number of reconstructed events (using the
reconstructed energy), and $N_{\rm th}$ is the number of thrown events
(using the generated energy).
Since the resulting TA SD spectrum is very consistent with
the HiRes spectrum assumed in the process, the bias due to this
choice of spectrum in the unfolding is negligible.  
The assumption of
a pure proton composition is based on the results of the HiRes experiment, 
and the systematic uncertainty connected with this choice is small for $\log_{10} (E/\unit{eV}) > 18.5$.

For the Auger spectra, the resolution effects are corrected using a
forward-folding method. The true flux description is based on a fit
to the measured Auger spectrum, assuming a combination of power laws
and a Fermi function-like suppression (e.g. Figure~5 of
\cite{auger}). The migration matrix is built from CORSIKA (QGSJet
II.3/Fluka) with GEANT4 detector simulations, assuming a 50:50 mixed
composition of proton and iron sample showers. The energy of the simulations is
scaled to take into account the known energy-scale discrepancy with
real data~\cite{Auger_hadronic}.  Distributions of shower observables (e.g. zenith angle,
stations per event, signal per station, $\chi^2$ of LDF fit, energy
resolution) are well matched by MC results.  The correction
contributes a systematic uncertainty to the flux of 5\%, calculated
assuming no energy re-scaling, and pure compositions. The systematic
uncertainty due to the choice of the {\it true} spectrum in the
simulation is negligible.

In the case of Yakutsk, no correction has yet been made to the
spectrum for the effects of resolution.


\subsection{ Budget of the Systematic Energy Uncertainty}
\label{section:sysuncert}

The energy scale for the Yakutsk SD spectrum is calibrated by
atmospheric Cherenkov light observations on a subset of showers.  As a
reminder, we point out that the Yakutsk SD energy parameter is either
$S_{300}$ for the trigger-500 array (the particle density 300\,m from
the core) or $S_{600}$ for the trigger-1000 array.  An attenuation
correction, derived from a constant intensity cut (CIC) method (and
cross-checked using Cherenkov measurements) is applied to convert
$S_{300}$ or $S_{600}$ to their equivalents for a vertical shower.
Finally, the primary particle energy is determined from
\begin{eqnarray*}
E_0 &=& (6.5 \pm 1.6) \times 10^{16} S_{300}(0^\circ)^{0.94±0.02}\, \unit{eV}\\
E_0 &=& (4.6 \pm 1.2) \times 10^{17} S_{600}(0^\circ)^{0.98±0.02}\, \unit{eV}
\end{eqnarray*}
where the constants are derived from the relationship between the
$S_{300}(0^\circ)$ or $S_{600}(0^\circ)$ and $Q_{400}$, the Cherenkov
light intensity at \unit[400]{m} from the core~\cite{Yakutskspectrum}.  The
systematic error in energy implied by these conversions is
approximately 25\%, attributed to calibration systematics in the
Cherenkov detectors and systematics in light transmission through the
atmosphere~\cite{Yakutskspectrum}.
A study has been done replacing the Cherenkov calibration by one that
converts $S_{300}$ or $S_{600}$ to energy via simulations
only~\cite{Dedenko}.  This results in $E_0 = (2.7 - 3.0) \times
10^{17} (S_{600})^{0.99}\,\unit{eV}$, giving energies that are only about 60\% of
those derived using the Cherenkov light measurements.  Interestingly,
this disagreement between shower simulations and a calorimetric
experimental method (Cherenkov light in this case) is in the opposite
direction to those found in the other experiments (see below).  In any
case, for the Yakutsk data discussed in the present paper, we assume
the original, Cherenkov light-based calibration from the original
spectrum publication~\cite{Yakutskspectrum}.

In the Auger SD analysis, the ground parameter used to extract the
primary energy is $S(1000)$, the water Cherenkov tank signal at
\unit[1000]{m} from the core.  Through an attenuation correction derived from
the CIC method, $S(1000)$ is converted to
$S_{38}$, the value of $S(1000)$ the shower would have if recorded at
the median zenith angle of $38^\circ$~\cite{auger_ecal_icrc11}.  This
parameter is then related to the primary energy using fluorescence
observations of a subset of showers, taking advantage of a
near-calorimetric fluorescence energy determination.  In these ways,
the energy assignment is nearly free of simulations, with the
exception being in the estimation of a small (of order 10\%, see
below) correction for {\it invisible}  energy, that part of the primary
energy carried into the ground by neutrinos and high-energy muons that
does not result in full fluorescence emission.

The Telescope Array SD analysis methods are broadly similar to that of
AGASA \cite{agasaexposure}, with the ground array energy parameter being $S(800)$, the scintillator
signal at 800\,m from the core.
TA uses simulations to determine the change in $S(800)$ as a function of
shower zenith angle at fixed energy. 
The first energy estimate $E_{\rm SD}$ from 
$S(800)$ is rescaled  by using the average FD-SD energy scale ratio obtained from
hybrid events, as $E = \left< E_{\rm FD}/E_{\rm SD} \right>_{\rm h} E_{\rm SD}$,
where $\left< E_{\rm FD}/E_{\rm SD} \right>_{\rm h} = 1/1.27$ \cite{TASD,TAenergyscale}.
The use of MC simulations is to account for any changes in the attenuation
function with energy, given that the CIC method is best applied at
lower energy where the statistical uncertainties are smaller.  On the
other hand, the simulation route is subject to uncertainties in both
the choice of hadronic model and the mass composition assumption.  
(The Auger collaboration has applied the CIC method with increasing cuts on
energy in an attempt to see any changes in the assumed attenuation
with zenith angle, but so far no significant change has been detected.)


\begin{table}[t] 
\begin{center}
\begin{tabular}{lccc}
\toprule
 & HiRes & Auger & TA  \\
\midrule
 Photometric calibration & 10\% & 9.5\% & 10\% \\
Fluorescence Yield & 6\% & 14\% & 11\% \\
Atmosphere & 5\% & 8\% & 11\% \\
Reconstruction & 15\% & 10\% & 10\% \\
Invisible Energy & 5\% & 4\% & incl. above \\
\midrule
{\bf TOTAL} & {\bf 17\%} & {\bf 22\%} & {\bf 21\%} \\
\toprule

\end{tabular} 
\caption{Estimates of contributions to systematic uncertainties in the fluorescence energy scale, for HiRes~\cite{HRSpectrum}, Auger~\cite{auger_ecal_icrc11} and the Telescope Array~\cite{TAHybrid}.  The total is the sum of the uncertainties in quadrature.}
\label{tab:sysuncert}
\end{center}
\end{table}

Auger and the Telescope Array both take great care in determining the
energy scale of fluorescence measurements, as this is the basis of the
energy measurements for both {\it hybrid} and SD spectra.  While the
fluorescence technique is conceptually elegant, with the amount of
light produced being directly proportional to the energy deposited by
the shower in the atmosphere, there are practical challenges.  Some of
these are expressed through estimates of the systematic
uncertainties related to the energy scale, listed in
Table~\ref{tab:sysuncert} for the two experiments and for HiRes.
Photometric calibration refers to the absolute calibration of the
telescopes and photomultipliers, and their wavelength response;
uncertainties in the fluorescence yield include those on the absolute
efficiency, its wavelength dependence, and its dependence on pressure,
temperature and humidity; atmospheric uncertainties include those
relating to Rayleigh and aerosol scattering; reconstruction
uncertainties are mainly related to the efficiency of light collection
in the telescope cameras; and the invisible energy uncertainties are
based on lack of knowledge of the true mass composition and on the
spread of predictions of invisible energy by different hadronic models.
The total systematic uncertainty on the fluorescence energy is of
order 20\% for the three experiments.

We will return to aspects of the fluorescence energy scale after
examining the level of agreement between the published energy spectra.

\begin{table}[!tb]
      \small
  \begin{center}
    \begin{tabular}{crrrr}
      \toprule      
      \multicolumn{4}{c}{Reference experiment} & Averaged reference\\
      \cmidrule(r){2-4}
      \small
      &  TA & Auger & Yakutsk & $\langle$Auger, TA$\rangle$\\
      \midrule
      TA  &-- & $0.0851 \pm 0.0027$  &$-0.2087 \pm 0.0056$& $0.0423 \pm 0.0027$ \\
\cite{TASD}& -- & $\chi^{2}$/ndof = 0.5060 & $\chi^{2}$/ndof =  0.5078&  \\
      & -- & P = 0.9874 & P = 0.9751& \\
      Auger &$-0.0843 \pm 0.0028$  & -- &$-0.2951 \pm 0.0048$ & $-0.0423 \pm 0.0027$ \\
\cite{auger} & $\chi^{2}$/ndof = 0.5671  & -- & $\chi^{2}$/ndof = 0.7478 &\\
      & P = 0.9586 & -- & P = 0.8227& \\
      Yakutsk  & $0.2110 \pm 0.0043$   &  $0.2985 \pm 0.0038$   & -- & $0.2562 \pm 0.0038$\\
\cite{Yakutskspectrum} & $\chi^{2}$/ndof = 0.6590 & $\chi^{2}$/ndof = 0.9550 & -- & \\
      & P = 0.9000 & P = 0.5342 & -- & \\
      HiRes I & -0.0009 $\pm$ 0.0063  & 0.0829 $\pm$ 0.0047&  -0.1978 $\pm$ 0.0103&  0.0406 $\pm$ 0.0047 \\
\cite{HRSpectrum}& $\chi^{2}$/ndof = 0.4604 &  $\chi^{2}$/ndof = 0.6530& $\chi^{2}$/ndof =0.5009 & \\
            & P = 0.9901 & P = 0.9144& P = 0.9855 & \\
      HiRes II & -0.0003 $\pm$ 0.0053& 0.0865 $\pm$ 0.0049&  -0.2093 $\pm$ 0.0033&  0.0442 $\pm$ 0.0049 \\
\cite{HRSpectrum}& $\chi^{2}$/ndof = 0.5135& $\chi^{2}$/ndof = 0.3497& $\chi^{2}$/ndof = 1.1036 & \\
            & P = 0.9784 & P = 0.9993& P = 0.3233 & \\
      \bottomrule
    \end{tabular}
    \caption{\label{tab:params} Parameters obtained from the
      energy-shift fitting for pairs of spectra.  Publication references for the spectra are given in the left-hand column.  We fitted for the
      logarithm of the scaling factors, with the entries representing
      $\log_{10}(\alpha)$ and its uncertainty. The value of the
      $\chi^{2}$/ndof and its probability are also given.  The columns
      indicate the reference spectrum, chosen from the currently operating experiments, including in the last column an
      average of the Auger and TA spectra (a convenient reference in the opinion of the WG).  The differences in the
      numbers for a given pair of spectra when switching the reference
      spectrum is due to the interpolations used in the method.}
  \end{center}
\end{table}

\section{Comparing Energy Scales}
\label{section:scales}

\begin{figure}[!t]
\begin{center}
\includegraphics[width=0.80\textwidth]{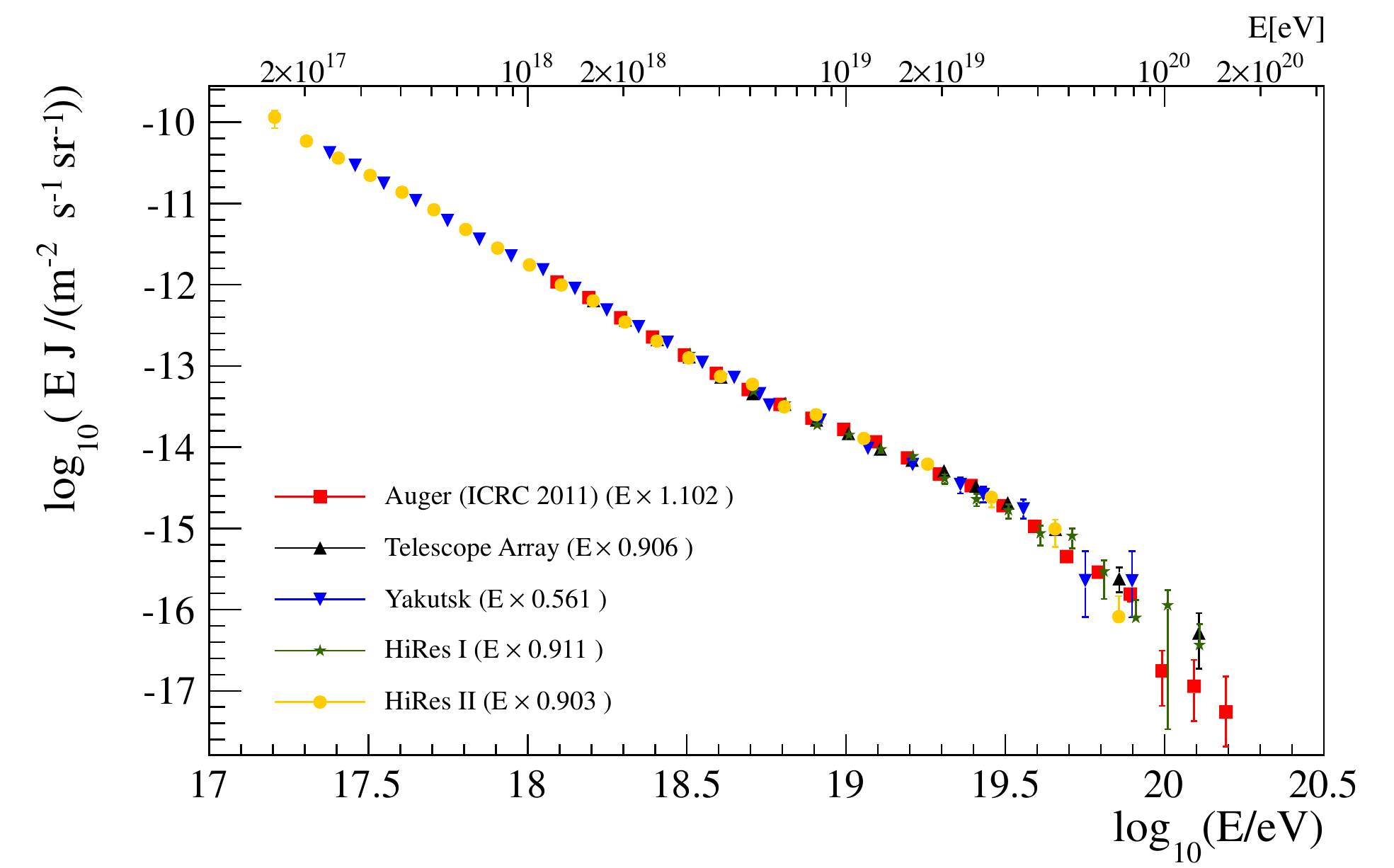}
\end{center}
\caption{Spectra from Figure~\ref{fig:current} after energy-rescaling.  The reference spectrum is the average of those from Auger and TA.  The values of  $\log_{10}\alpha$ are given in Table~\ref{tab:params}, and the values of $\alpha$ are indicated on the plot.}
\label{fig:energySpectraRescaled1}
\end{figure}

\begin{figure}[!tb]
\begin{center}
\includegraphics[width=0.80\textwidth]{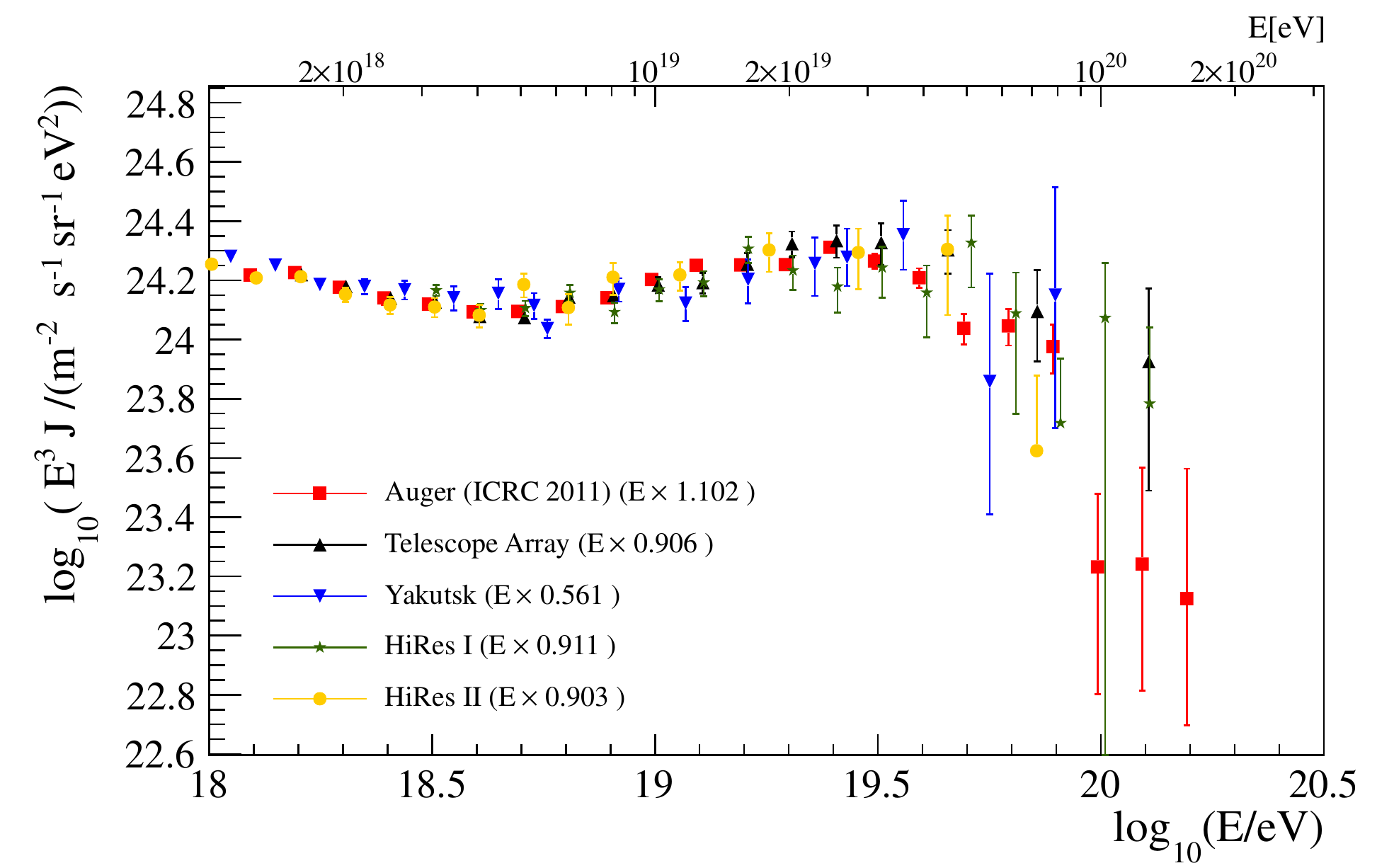}
\end{center}
\caption{Re-scaled spectra from Figure~\ref{fig:energySpectraRescaled1}, but in
the form $E^3 J$}
\label{fig:energySpectraRescaled2}

\begin{center}
\includegraphics[scale=0.5]{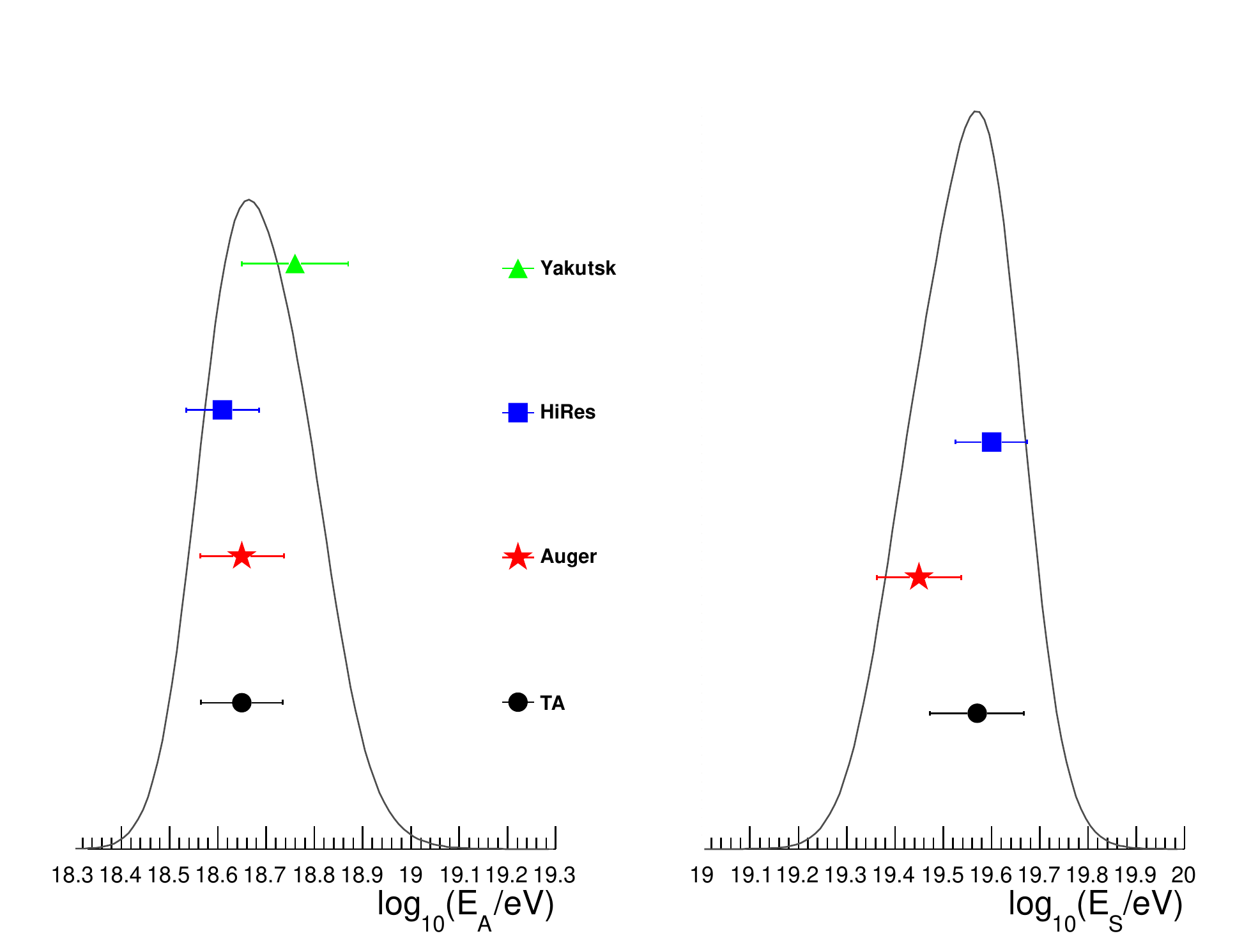}
\end{center}
\caption{Break-point energies for the triple-power law fits after
  energy rescaling, to be compared with the original positions in
  Figure~\ref{fig:BreakPoints0}.  The error bars again represent the
  statistical uncertainty folded with the systematic uncertainty in
  the energy scale for each experiment.  The reference spectrum is the
  average of Auger and TA.}
\label{fig:BreakPoints1}
\end{figure}

The WG undertook an exercise to see if the various spectra
could be brought into better agreement through a simple scaling of the
energy scale.  This assumes that any current disagreement is based solely on
the energy scale, and not on other factors such as aperture
calculation or the treatment of energy resolution, but we believe that
the results are informative.
As input to the calculation we took energy spectra published by the
Yakutsk, HiRes, Telescope Array and Auger collaborations.  Then, to obtain the energy
normalizations for the individual energy spectra, we performed a minimization of
the following $\chi^2$ function to compare one spectrum (called the
{\it reference} spectrum)
to one from another experiment,
\begin{equation}
  \chi^2 = \sum_{i=0}^{i < N_0}
  \frac{\left(F_{0,i}-\hat{F}_{k}(x_{0,i}^{0}-\log_{10}\alpha_{k}
    )\right)^2}{\sigma_{ 0,i}^2 +
    \hat\sigma^2(x_{0,i}-\log_{10}\alpha_{k})} + 
  \frac{\left(1.-\alpha_{k}\right)^2}{\sigma_{sys,
      k}^2+\sigma_{sys,0}^2}
\end{equation}
where the parameters are defined below,
\begin{itemize}
\item $k$ is the index of the experiment, $k\in\{0,1,2,3\}$, where
  $k=0$ represents the energy spectrum of the {\it reference} experiment
\item $N_{k}$ is the number of data points for the $k$-th spectrum, and $i$ is the index for the data points, $i\in\{0,\ldots,N_{k}-1\}$
\item $E_{k,i}$ is the  energy and $J_{k,i}$ is the differential flux of the $i$-th data point
\item $(F_{k,i}, x_{k,i})$ are the data points for the fit, with
  $x_{k,i}=\log_{10}(E_{k,i})$ and $F_{k,i}= E_{k,i} J_{k,i}$
\item $\sigma_{k,i}$ is the statistical uncertainty of $F_{k,i}$
\item $\alpha_{k\in\{0,\ldots,3\}}$ are the multiplicative scaling factors for
  the energy.  We fit for $\log_{10}(\alpha_{k\in(0,3)})$ since this
  factor is symmetric when comparing the shifts of two spectra against
  each other.
\item $\hat F$, $\hat \sigma$ is the flux multiplied by energy, and
  its uncertainty, estimated for $k>0$ from a linear interpolation in
  log-log space between neighboring bins
\item $\sigma_{sys, k}$ represent the energy systematic uncertainties,
  taken as 22\% for the Telescope Array and Pierre Auger measurements, 25\% for the Yakutsk measurement and 17\% for the HiRes experiment.
\end{itemize}
For differential spectra $J(E)$, we compared $F=E\,J(E)$ from the
different experiments, because $E\,J(E) \propto E\,dN/dE \propto dN/d(\log_{10}
E)$ is what is actually measured by the experiments.  Because there is
an assumption in this $\chi^2$ minimization that the uncertainties on
the variables are normally distributed, we perform the fit only up to
$\log_{10} (E/{\rm eV}) = 19.5$ to avoid the Poisson statistics of low
event counts.  The spectra investigated are listed in  Table~\ref{tab:params}
with their publication references, together with the results of this scaling \
exercise.  Each column is for a particular
reference spectrum, and the fitting was done in each case for just the
two spectra indicated by the column and the row.  For each fit, the
numbers represent $\log_{10}\alpha$ and its uncertainty.  The reference spectra
were taken to be those of the currently operating experiments, and
the final
column uses an average of the Auger and TA spectra as the reference, a 
convenient baseline defined by the WG.
The rescaled energy spectra, referenced to the average of TA and
Auger, are illustrated in Figures~\ref{fig:energySpectraRescaled1}~\&~\ref{fig:energySpectraRescaled2}.

Examining
Figures~\ref{fig:energySpectraRescaled1}~\&~\ref{fig:energySpectraRescaled2}
and the $\chi^2$ values in the body of Table~\ref{tab:params}, we see
that the rescaled spectra are in very good agreement in both shape and
normalization.  We searched for evidence of energy dependent scaling
factors of the form $\alpha (E/{\rm EeV})^\beta$ and found values of $\beta$
consistent with zero.  With the energy scaling, it is not surprising that
the positions of the spectral features are now more consistent, as shown in 
Figure~\ref{fig:BreakPoints1}.

Overall we find that the energy scales of the TA and HiRes experiments
are nearly identical, and that TA and Auger have a an energy scale
difference expressed as $\log_{10}\alpha=\pm 0.085$, or that energies
are reconstructed by Auger at values $\sim 20$\% smaller than TA for
the same flux.  Such a difference is entirely consistent with the
systematic energy uncertainties listed in Table~\ref{tab:sysuncert}, a
topic we will return to in the next section.  We also find that
Yakutsk energies are overestimated by 60\% with respect to TA and 95\%
with respect to the Auger scale.  These inconsistencies are more
difficult to explain in terms of estimated systematic uncertainties,
which are estimated to be 25\% for the Yakutsk Cherenkov light
calibration.  We note that the alternative Yakutsk calibration
procedure based on shower simulations described earlier
(Section~\ref{section:sysuncert}) would be more consistent with the
fluorescence energy determinations.

This statement should not be taken to suggest that air shower
simulations are superior to calibrating with calorimetric methods such
as Cherenkov or fluorescence light.  There are large 
uncertainties on the hadronic interaction model, and in calculations done by the TA and Auger
experiments, inconsistencies are also seen between a  {\it simulation} energy
scale and that derived from fluorescence measurements.  The size of
the discrepancy is dependent on the hadronic model and the mass
composition assumptions.  In TA, simulations of proton showers using
QGSJET-II predict a energy for a given $S(800)$ which is 27\% lower
than the energy derived from fluorescence
observations~\cite{TASD,TAenergyscale}.  Auger also sees an effect in the same
direction, with simulations (protons or iron) underestimating the
ground signal for a given (fluorescence-derived)
energy~\cite{Auger_hadronic}.  Note that the simulation/fluorescence
energy difference seen by Auger and TA is in the {\it opposite}
direction to that seen in the simulation/Cherenkov comparison of
Yakutsk.

\begin{figure}[!bt]
\begin{center}
\includegraphics[scale=0.7]{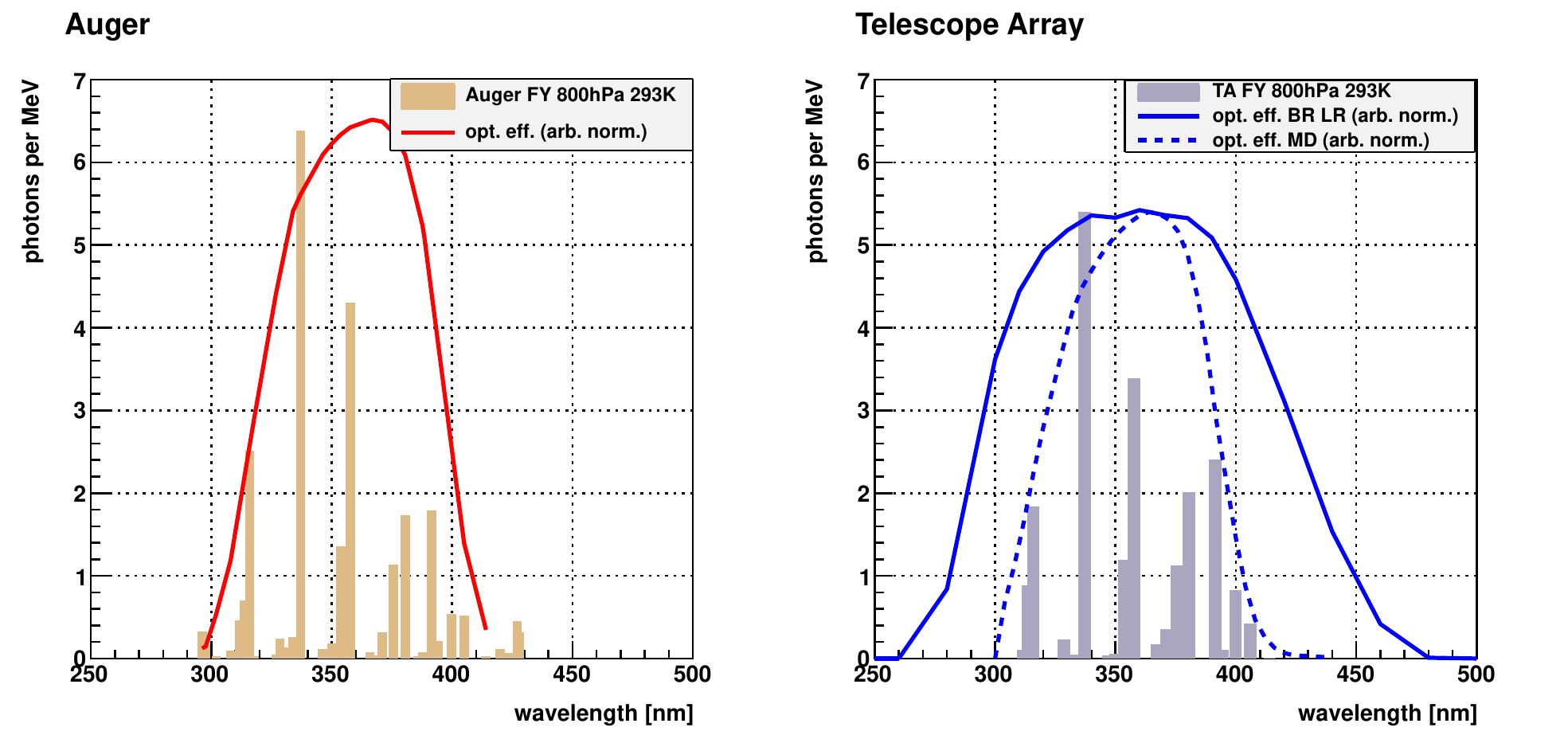}
\end{center}
\caption{Optical efficiency curves (arbitrary normalization) together
  with the assumed fluorescence spectra for Auger and the Telescope Array.
  The spectra are for dry air at a pressure of 800\,hPa and a
  temperature of 293\,K.  In the case of the Telescope Array, there
  are different optical efficiency curves for the Black Rock and Long
  Ridge sites on one hand, and the former HiRes telescopes now at the
  Middle Drum site on the other. }
\label{fig:efficiency}
\end{figure}

\subsection{Analysis \& Calibration Differences}

The comparisons of spectra in the previous section suggest that a
simple rescaling of energy can bring the results into agreement.  In
the case of the Yakutsk/TA and Yakutsk/Auger comparisons the required
rescaling is somewhat outside that allowed by the known systematic
uncertainties, but Auger/TA rescaling is perfectly consistent with the
22\% and 21\% energy scale systematics of Auger and TA respectively
(Table~\ref{tab:sysuncert}).  In this section we concentrate on some
differences in the analysis methods of Auger and TA relevant to the
energy scale.

\subsubsection{Fluorescence Yield}

The two experiments currently use different laboratory measurements of
the fluorescence yield, its wavelength dependence, and its dependence
on pressure, temperature and humidity.  The Auger approach is outlined
in \cite{auger_ecal_icrc11}; it can be summarized as using the 2004
measurements of Nagano et al. for the yield of the 337\,nm band, with the
relative wavelength spectrum and its dependence on pressure,
temperature and humidity coming from the AIRFLY collaboration.  In
contrast, TA uses spectral lines and their relative intensities from
the FLASH experiment~\cite{FLASH}, but rescales these so that the
total yield between 300-400\,nm is consistent with
\cite{Kakimoto}, see \cite{TAenergyscale}. Pressure and temperature corrections are also taken
from \cite{Kakimoto}, and no humidity correction is currently applied.

The telescopes of Auger and TA have a particular spectral response, and
this must be taken into account when attempting to understand the
effect of different fluorescence yield models.  We refer to the
spectral response of a telescope as its ``optical efficiency'',
defined as the product of mirror reflectivity, optical filter
transmission and PMT quantum efficiency as a function of wavelength.
In the case of TA this also includes the ``paraglas'' window
transmission, and in the case of Auger it includes the transmission of
the corrector ring optics.  Figure~\ref{fig:efficiency} shows the
relative optical efficiency curves for Auger and TA, 
together with the assumed fluorescence spectra for the two
experiments, for a pressure of 800\,hPa and a temperature of 293\,K.
The TA experiment includes former HiRes telescopes at the Middle Drum
site, so two optical efficiency curves are shown.  The optical
efficiency for the TA Long Ridge and Black Rock sites has a wider
bandpass than either the Auger or the Middle Drum optical
efficiencies.  Such a bandpass increases the signal available, but
also increases the level of night sky background.  The two
collaborations have decided to optimize their sensitivity in different
ways.

Given the diversity in laboratory measurements of the fluorescence
spectrum and normalization, it would be fair to conclude that we do
not yet know the {\it true} fluorescence description, though progress is being made on this front~\cite{yield}.  However, we do
have two possible versions of it in the models applied by Auger and
TA.  As an exercise, the WG considered the following question: `` What
would be the effect on one experiment's energy scale if the
fluorescence model assumed by the {\it other} experiment was actually
the ``true'' one?''.  To do this we first ignored the effect of
atmospheric scattering in changing the shape of the fluorescence
spectrum, and calculated the following ``signal'',
$$S = \sum_\lambda Y(\lambda) \epsilon(\lambda)$$ where $Y(\lambda)$
is the fluorescence spectrum and $\epsilon(\lambda)$ is the optical
efficiency, both taken from Figure~\ref{fig:efficiency}, with
$\epsilon(\lambda)$ normalized to 1 at 375\,nm.  As an example,
the value of $S$ for the Auger spectrum and efficiency is 18.1
``signal'' units.  For a given experiment, we fixed the $\epsilon$
curve, but were able to switch the $Y$ curve.  

First, we consider the possibility that the ``true'' fluorescence model
is the TA description.  Then for a shower of some energy $E_0$, Auger
would record a signal of $S=\sum Y_{\rm TA}\epsilon_A=16.1$\,units, while
it would expect, on the basis of its assumed fluorescence model, to
receive $S=\sum Y_{\rm A}\epsilon_A=18.1$\, units.  Here, the subscripts
on $Y$ or $\epsilon$ refer to either ``TA'' or ``A'' for Auger.  As a
result, Auger will reconstruct an energy which is too {\it low} by
$1-(16.1/18.1) = 11\%$  or $E_{\rm TA}/E_{\rm A} = 1.12$.

Now consider that the ``true'' yield description is the Auger model.
In this case, TA will receive from a shower of some energy a signal of
$S=\sum Y_{\rm A}\epsilon_{\rm TA}=22.7$\,units, while expecting a signal of
$S=\sum Y_{\rm TA}\epsilon_{\rm TA}=19.4$\,units.  As a result TA will
reconstruct an energy which is too {\it high} by $(22.7/19.4)-1 =
17\%$, or $E_{\rm TA}/E_{\rm A} = 1.17$.  (Here we used the Black Rock and
Long Ridge optical efficiency.  When using the Middle Drum (HiRes)
efficiency, we get a similar result of $E_{\rm TA}/E_{\rm A} = 1.14$).

The actual numbers depend on which (if either) of the fluorescence
models is the true one.  But note that in both cases, the Auger
reconstructed energy is lower than the TA reconstructed energy, in
agreement with the direction of the discrepancy in the energy spectra.
However, so far we have ignored the effect of the atmosphere in two
ways; first the effect of atmospheric transmission, especially
Rayleigh scattering, which will preferentially attenuate shorter
wavelengths; and secondly the effect of humidity quenching of
fluorescence, taken into account by Auger but not yet by TA.  The
first effect was tested by reconstructing a sample of real Auger air
showers ($\log_{10} (E/{\rm eV}) > 18.5$) using the TA fluorescence description,
properly taking into account the differential attenuation across the
fluorescence spectrum.  The result derived above ignoring attenuation,
$E_{\rm TA}/E_{\rm A} = 1.12$ was reduced to $E_{\rm TA}/E_{\rm A} = 1.08$
here.  The effect of average levels of humidity is to reduce the yield
by 5\% and hence increase the reconstructed energies by
5\%~\cite{auger_atmosphere}.  Because Auger applies this correction
and TA doesn't, the expected gap between TA and Auger energies is
further reduced to $E_{\rm TA}/E_{\rm A} = 1.03$.  All these figures assume
that the true fluorescence model is the TA one.

To summarize, if the true yield model is the TA one, TA and TA (Middle
Drum) would reconstruct the correct energy, and Auger's energy would
be low by about 3\%, ignoring all other effects.  If the true yield
model is the Auger one, Auger's energies would be correct, and TA
energies would be too high by 8\%, and TA (Middle Drum) energies too
high by 5\%.  Similar results have been determined by Vazquez et
al.~\cite{Arqueros}.

\subsubsection{Atmospheric Monitoring}

The atmosphere attenuates fluorescence and Cherenkov light primarily
via Rayleigh and aerosol scattering, and via cloud cover.  The density
and temperature profile of the atmosphere determines the fluorescence
yield, and the density profile governs the conversion of height to atmospheric
depth.  Each of the fluorescence observatories has a major program of
atmospheric monitoring, outlined in Table~\ref{tab:atmosphere}.  The
molecular profile of the atmosphere may be determined by radiosonde
launches or by data assimilation products like GDAS~\cite{GDAS}.
Aerosol conditions may be characterised by traditional LIDAR systems
or by a bistatic LIDAR system employing a Central Laser Facility
(CLF).  Currently, the Auger Observatory uses molecular profiles from
GDAS with 3-hour latency, hourly measurements of aerosols with its CLF
and XLF, and sub-hourly measurements of cloud cover with LIDARs and
infra-red cameras \cite{auger_atmosphere,CLF}.  The Telescope Array uses publicly acquired
radiosonde data, a fixed aerosol model derived from LIDAR
observations, and infra-red cameras for cloud cover \cite{ta_atmosphere}.  Hourly aerosol
measurements will be used in the future as the CLF analysis matures.

\begin{table}[!tb]
\begin{center}
\begin{tabular}{cccccc}
\toprule
  &  Radiosonde & LIDAR & CLF & IR Camera & Other\\
\midrule
 &Stereo: public data,&N/A &Stereo:hourly&N/A& \\
HiRes&Salt Lake City \&& &Mono: mean& & \\
(5\%)&Elko, 200km from& &aerosol model.& & \\
\cite{hires_atmosphere}&site. Mono: US& & & & \\
 &standard atmos.& & & & \\
\midrule
&Monthly models from&4 sites, hourly. Cloud&2 sites. Hourly&4 sites, 5 min&aerosol\\
Auger&own radiosonde data.&cover \& height, cross&VAOD. Used &scans. Data&phase funct.\\
 (8\%)&Replaced by GDAS &check of VAOD.&in analysis.&selection via&\& $\lambda$ depend.\\
\cite{auger_atmosphere,GDAS}&with 3hr update.&     &    &LIDAR. &monitors.\\
\midrule
&Public data. Salt&One site (BR) twice &Hourly, will&1 site, hourly & \\
TA&Lake City \& Elko,&a day. Use fixed&use in analysis&monitoring.& \\
(11\%)&200\,km from site&model in recon&in the future &Used for data& \\
\cite{ta_atmosphere}&                &(average VAOD,& &selection.& \\
       &                &scale height, HAL)& & & \\
\bottomrule
\end{tabular}
\caption{Some highlights of the atmospheric monitoring activities at
  the former HiRes site, and at the Auger Observatory and the
  Telescope Array.  The percentages in the left-hand column refer to
  the estimated systematic uncertainties in energy attributed to the
  atmosphere.  References are also given in the left-hand column.  CLF
  refers to ``Central Laser Facility''.}
\label{tab:atmosphere}
\end{center}
\end{table}

\subsubsection{Photometric Calibration}

Auger and the Telescope Array use different techniques for photometric
calibration of their telescopes, and both nominate a significant
$\sim10\%$ systematic uncertainty in energy due to calibration.
Details of the calibration methods are given in
\cite{auger_calibration} for Auger and \cite{ta_calibration} for TA.
The TA experiment also has an opportunity for an end-to-end
calibration with a electron linear accelerator system~\cite{linac}.
The WG sees an opportunity to inter-calibrate the two experiments in
some way, perhaps using a flying calibrated light source (see below).

\subsubsection{Invisible Energy Corrections}

\begin{figure}
\begin{center}
\includegraphics[scale=0.5]{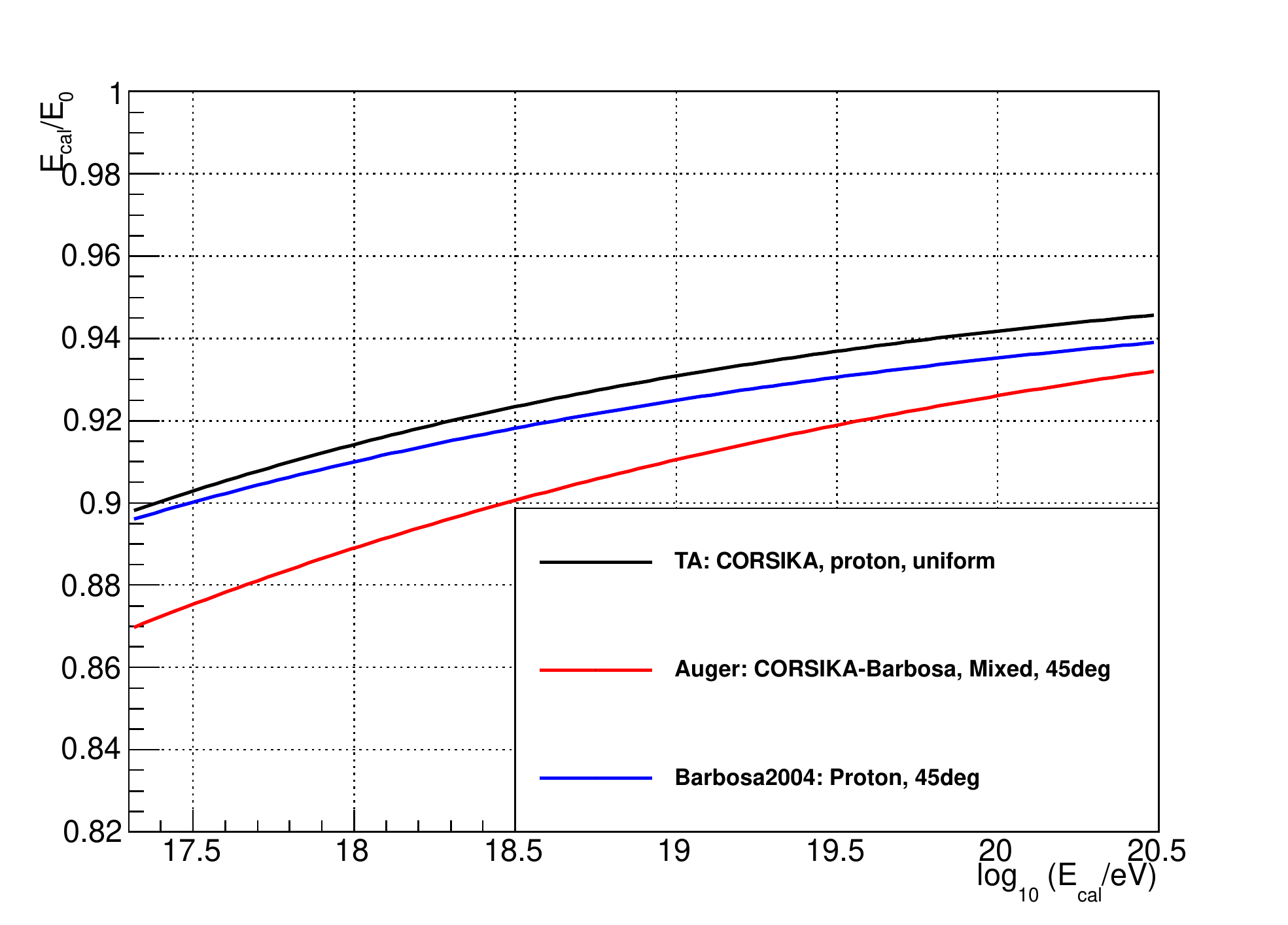}
\end{center}
\caption{Comparison of the conversion factors from calorimetric energies to primary energies.  See text for details.}
\label{fig:MEC}
\end{figure}

As discussed in Section~\ref{section:sysuncert}, a correction must be
applied to the calorimetric energy $E_{\rm cal}$ measured by a
fluorescence detector to account for invisible energy and to derive
the primary energy $E_0$.  The form of the correction is the same for
Auger and TA, $E_{\rm cal}/E_0 = A - B (E_{\rm cal}/{\rm EeV})^{-C}$,
but the constants $A,B,C$ are derived from different sources.  The
Auger values come from \cite{Barbosa} assuming a QGSJET01 hadronic
model, a mixed composition of 50:50 protons and iron, and a fixed
$\theta = 45^\circ$ zenith angle, resulting in $A = 0.967, B = 0.078$ and $C =
0.140$.  The Telescope Array group has performed their own CORSIKA
simulations for
a pure proton composition with the QGSJET-II model, and
uniform arrival directions ($\theta < 60^{\circ}$) to extract $A = 0.963, B = 0.049$ and $C =
0.181$.  These curves are plotted in Figure~\ref{fig:MEC}.  The
separation between the Auger and TA correction is never larger than
2\%, and is in the opposite direction to that implied by the energy
spectrum differences discussed in Section~\ref{section:scales}.  In any case,
the difference is well within the systematic uncertainty assigned to
this correction by both experiments (Table~\ref{tab:sysuncert}).

\section{Future Directions}

In the future, each experiment will strive to reduce systematic
uncertainties in energy calibration, attenuation curves, and
reconstruction issues, and continue to employ cross-checks to validate
these methods.  Examples of cross-checks include using remote laser
checks of photometric calibration, and hybrid data checks of surface
signal attenuation with zenith angle.  We will all continue to improve
the important atmospheric monitoring and corrections.

As a community, we must maintain contacts between experiments. It may
be necessary to have a high-level organization to encourage this,
especially if more than a yearly meeting is desired.  The WG supports
the adoption of a world-wide fluorescence yield model, including
humidity, pressure and temperature dependencies~\cite{yield}.  Where possible, we
should use a common set of procedures. If some procedure is not the
preferred one for a particular experiment, we suggest that it can be
employed as a cross-check against the preferred method. For example TA
could try to use the CIC method to cross-check the MC-derived SD
attenuation curve, and Auger could do the reverse.  We should attempt
to cross-check the TA and Auger FD photometric calibration, perhaps
employing a calibrated light source flown by an octocopter, or by
using a standard roving laser system. 
Finally, we can learn from each other
about the best methods of atmospheric characterization.

\section{Conclusions}

The working group has examined the energy spectra and techniques of
the currently operating experiments, the Pierre Auger Observatory, the
Telescope Array, and Yakutsk, with some reference to previous
measurements by AGASA and HiRes.  We find that the spectra are in
agreement after making energy scale shifts consistent with the quoted
energy systematic uncertainties, with the possible exception of
Yakutsk.  After energy rescaling, the spectral normalizations and
shapes are very consistent, and the positions of the spectral ankle and
the suppression are in agreement. 

The existence of the spectral ankle is certainly supported by all
experiments.  As is well known, the high-energy spectral suppression
is not compatible with the AGASA spectrum, and the Yakutsk spectrum
currently has insufficient exposure to properly detect it.  But HiRes, TA and Auger all
see the suppression with good statistical significance, with HiRes
associating it with the GZK cut-off~\cite{GZK} because of their
measurement of protonic composition at the highest energies, and the
suppression's position in energy~\cite{HRSpectrum}.  The Auger Observatory's measurement
of mass composition at the highest energies is not consistent with protons
under current hadronic interaction models~\cite{massWG}, and so that collaboration has not
yet associated the suppression with a particular mechanism.

We have started to investigate the alternative methods employed by the
experiments.  Essentially all analysis differences are based on well
founded preferences, and should lead to equivalent outcomes. This
should be tested in the future by cross-checks of methods using those
alternatives.  The energy scale difference seen between Auger and TA
is consistent with systematic uncertainties, but does deserve further
study.  Common models (e.g. fluorescence yield) should be used where
possible, and cross-checks of photometric calibrations and
atmospheric corrections could be very instructive.

The members of the WG very much enjoyed their interactions which were
played out in a friendly and constructive way.  We hope that this
process will be continued into the future to benefit all studies of
ultra-high energy cosmic rays.


\begin{thebibliography}{}

\bibitem{auger} F.\, Salamida {\it et al.} \pao, {\it Proc. 32nd Int. Conf. Cosmic Rays},
\textbf{2} (2011) 145.
\bibitem{TASD} D. Ivanov, B.T. Stokes, G.B. Thomson {\it et al.} \tac,
  {\it Proc. 32nd Int. Conf. Cosmic Rays}, \textbf{2} (2011) 258., T. Abu-Zayyad {\it et al.}. arXiv:1205:5067 (submitted to {\it Phys. Rev. Lett.})
\bibitem{TAHybrid}D. Ikeda {\it et al.} \tac, {\it Proc. 32nd Int. Conf. Cosmic Rays}, \textbf{2} (2011) 238.
\bibitem{Yakutskspectrum} Egorova V.P.,Glushkov A.V.,Knurenko S.P., {\it et al.}, {\it Nuclear Physics B (Proc.Suppl.)} \textbf{136C} (2004) 3; Ivanov A.A., Knurenko S.P., Pravdin M.I. and Sleptsov I.E., {\it Moscow University Physics Bulletin}, 2010, \textbf{65}, No 4 (2010) 292 - 299.
\bibitem{agasaexposure} K. Shinozaki \& M. Teshima, {\it Nuclear Physics B (Proc. Suppl.)} \textbf{136} (2004) 18.
\bibitem{hiresIexposure}  http://www.physics.rutgers.edu/$\sim$dbergman/HiRes-Monocular-Spectra-200702.html

\bibitem{Unger} M.\,Unger, ``EAS Studies of Cosmic Rays above
  $10^{16}$eV'', {\it Proc. 32nd Int. Conf. Cosmic Rays}, \textbf{12}
  (2011).

\bibitem{HRSpectrum} R.U. Abbasi {\it et al.} [Hires Collaboration], {\it Phys. Rev. Lett.} \textbf{100} (2008) 101101. 
\bibitem{agasaspectrum} M. Takeda {\it et al.}, {\it Phys. Rev. Lett.} \textbf{81} (1998) 1163.

\bibitem{Dedenko} L.G. Dedenko, D.A. Podgrudkov and
  T.M. Roganova, {\it Physics of Atomic Nuclei} \textbf{70} No. 10 (2007)
  1759-1763

\bibitem{auger_ecal_icrc11} R.\.Pesce {\it et al.} \pao, {\it Proc. 32nd
  Int. Conf. Cosmic Rays}, \textbf{2} (2011) 214.

\bibitem{TAenergyscale} Y.~Tsunesada {\it et al.}, \tac,
  {\it Proc. 32nd Int. Conf. Cosmic Rays}, \textbf{2} (2011) 1270

\bibitem{Auger_hadronic}J.\.Allen {\it et al.} \pao, {\it Proc. 32nd
  Int. Conf. Cosmic Rays}, \textbf{2} (2011) 83.

\bibitem{FLASH} R.\,Abbasi {\it et al.}, {\it Astropart. Phys.}, \textbf{29} (2008) 77-86.

\bibitem{Kakimoto} F.\,Kakimoto {\it et al.} {\it Nucl. Instr. Meth.} \textbf{A 372} (1996) 527-533.

\bibitem{yield} B.\, Keilhauer {\it et al.}, ``Nitrogen fluorescence in air for observing extensive air showers'', this conference (2012).

\bibitem{auger_atmosphere} J.\,Abraham {\it et al.} \pao, {\it Astropart. Phys.},
  \textbf{33} (2010) 108-129.

\bibitem{CLF} B.\,Fick {\it et al.}, {\it JINST},
  \textbf{1} (2006) 11003.

\bibitem{Arqueros} J.R.\,Vazquez {\it et al.}, ``The impact of the
  fluorescence yield on the energy reconstruction of UHECR showers'',
  this conference (2012).

\bibitem{hires_atmosphere} R.\,Abbasi {\it et al.},
  Astropart. Phys. \textbf{25} (2006) 74.

\bibitem{GDAS} P.\,Abreu {\it et al.} \pao, Astropart. Phys.,
  \textbf{35} (2012) 591-607.

\bibitem{ta_atmosphere} T.~Tomida {\it et al.},  {\it Nucl. Instr. Meth.} \textbf{A 654} (2011) 653-660, Y.~Takahashi {\it et al.}, {\it UHECR2010}, {\it AIP Conf. Proc.} \textbf{1367}, 157 (2011), F.~Shibata {\it et al.}, {\it UHECR2010}, {\it AIP Conf. Proc.} \textbf{1367}, 165 (2011), Y.~Kobayashi {\it et al.}, {\it UHECR2010}, {\it AIP Conf. Proc.} \textbf{1367}, 169 (2011)

\bibitem{auger_calibration} J.\,Abraham {\it et al.} \pao,  {\it Nucl. Instr. Meth.} \textbf{A 620} (2010) 227-251.

\bibitem{ta_calibration} S.~Kawana {\it et al.}, {\it Nucl. Instr. Meth.} \textbf{A 681} (2012) 68-77, H.~Tokuno {\it et al.} {\it Nucl. Instr. Meth.} \textbf{A 601} (2009) 364-371

\bibitem{linac} T.~Shibata et al., Proc. 31st Int. Conf. Cosmic Rays (2009);
T.~Shibata {\it et al.}, {\it Nucl. Instr. Meth.} \textbf{A 597} (2008) 61., T.~Shibata {\it et al.}, {\it UHECR2010}, {\it AIP Conf. Proc.} \textbf{1367}, 44 (2011),

\bibitem{Barbosa} H.~Barbosa {\it et al.}, {\it Astropart. Phys.} \textbf{22} (2004) 159.

\bibitem{GZK} K.~Greisen, {\it Phys. Rev. Lett.} \textbf{16} (1966) 748; G.T. Zatsepin and V.A. Kuz'min, Pis'ma Zh. Eksp. Teor. Fiz \textbf{4} (1966) 114 [JETP Lett. \textbf{4} (1966) 78].

\bibitem{massWG} E.~Barcikowski {\it et al.}, ``Mass Composition Working
  Group Report'', this conference (2012).


\end{thebibliography}
\end{document}